\documentclass[12pt,preprint,prd,showpacs,showkeys]{revtex4}
\usepackage{amsmath,amssymb}
\usepackage{graphicx}

\begin{document}
\title{Gamma-Rays Produced in Cosmic-Ray Interactions and the TeV-band Spectrum of RX 
J1713-3946}
\author{C.-Y. Huang\footnote{Corresponding author: Tel.: +1-515-2945062; fax: +1-515-2946027\\
                        {\it E-mail address:} huangc@iastate.edu (C.-Y. Huang)}}
\author{S.-E. Park}
\author{M. Pohl}
\author{C. D. Daniels}
\affiliation{Department of Physics and Astronomy, Iowa State University, Ames, IA 50011}

\def\AstroPartPhys#1#2#3{Astropart. Phys.~#1~(#2)~#3.}
\def\AA#1#2#3{Astron. Astrophys.~#1~(#2)~#3.}
\def\APSS#1#2#3{Astrophys. \& Space Science~#1~(#2)~#3.}
\def\APJ#1#2#3{Astrophys. J.~#1~(#2)~#3.}
\def\MNRAS#1#2#3{Mon. Not. R. Astron. Soc.~#1~(#2)~#3.}
\def\MNRASLett#1#2#3{Mon. Not. R. Astron. Soc. Lett.~#1~(#2)~#3.}
\def\PRD#1#2#3{Phys. Rev. D~#1~(#2)~#3.}
\def\PRL#1#2#3{Phys. Rev. Lett.~#1~(#2)~#3.}
\def\PLB#1#2#3{Phys. Lett. B~#1~(#2)~#3.}
\def\NPB#1#2#3{Nucl. Phys. B~#1~(#2)~#3.}
\def\JPhysG#1#2#3{J. Phys. G.~#1,~(#2)~#3.}
\def\Nature#1#2#3{Nature~#1~(#2)~#3.}
\def\PR#1#2#3{Phys. Rev.~#1~(#2)~#3.}
\def\PASJ#1#2#3{PASJ~#1~(#2)~#3.}
\def\PhysRept#1#2#3{Phys. Rept~#1~(#2)~#3.}

\def\mkp{\bf }
\def\huangc{\bf}

\begin{abstract}
In this work we study the individual contribution to diffuse $\gamma$-ray emission from the 
secondary products in hadronic interactions generated by cosmic rays (CRs), in addition to the 
contribution of $\pi^0$ decay via the decay mode $\pi^0 \rightarrow 2\gamma$. For that
purpose we employ the Monte Carlo particle collision code DPMJET3.04 to determine
the multiplicity spectra of various secondary particles with $\gamma$'s as the final decay state, that result
from inelastic collisions between cosmic-ray protons and Helium nuclei and the interstellar medium
with standard composition. By combining the simulation results with a parametric model of $\gamma$-ray
production by cosmic rays with energies below a few GeV, where DPMJET appears unreliable, we thus derive
an easy-to-use $\gamma$-ray production matrix for cosmic rays with energies up to about 10 PeV.

We apply the $\gamma$-ray production matrix to the GeV excess in diffuse galactic $\gamma$-rays
observed by EGRET. Although the non-$\pi^0$ decay components have contributed to the total emission with a different spectrum 
from the $\pi^0$-decay component, they are insufficient to explain the GeV excess.

We also test the hypothesis that the TeV-band $\gamma$-ray emission of the shell-type SNR RX J1713-3946 observed with HESS is caused by shock-accelerated hadronic 
cosmic rays. This scenario implies a very high efficacy of particle acceleration, so the particle spectrum is expected to continuously harden toward high energies on 
account of cosmic-ray modification of the shock. 
Using the $\chi^2$ statistics we find that a continuously softening spectrum is strongly preferred, in contrast to expectations. A hardening spectrum has 
about 1\% probability to explain the HESS data, but then only if a hard cut-off at 50-100 TeV is imposed on the particle spectrum.
\end{abstract}
\maketitle
\section{Introduction}\label{Section:Introduction}
It is generally believed that diffuse galactic $\gamma$-rays are a good probe of 
the production sites and also the propagation of accelerated charged particles in the 
Galactic plane (see \cite{Nishimura80,Erlykin98} and references therein). 
Complementary knowledge of the cosmic-ray (CR) propagation is based on the isotopic 
composition and the content of secondary particles/anti-particles of the locally 
observed CRs which, however, requires a number of model-dependent assumptions. 
Also, it is not clear whether or not the locally observed CR spectra can be taken 
as representative for the entire Galaxy \cite{Cowsik79,Pohl98,Buesching05}. 

The observed diffuse Galactic $\gamma$-ray emission is composed of three spectral
components \cite{Hunter97,Aharonian00,Dermer86APJ,Stephens81,Mori97}: 
$\gamma$-rays from bremsstrahlung and from inverse Compton (IC) scattering, 
and also from the decay of secondary particles, mostly neutral pions, produced in collisions of 
CR nuclei with ambient gas. However, the relative contributions of the first two processes with 
respect to the $\pi^0$ decays is uncertain \cite{Pohl02}, thus leading
to a considerable uncertainty
in a physical interpretation of the observed Galactic $\gamma$-ray emission \cite{Strong04}.

The situation is similarly difficult for individual sources of GeV-to-TeV $\gamma$-rays, which 
may have been produced in hadronic interactions or leptonic processes like inverse Compton scattering. 
The most notable examples of sources difficult to interpret are 
AGN jets \cite{Pohl00,Muecke01,Muecke03,Boettcher04} and recently observed shell-type SNR's 
\cite{Aharonian04Nature,Aharonian05AA,Aharonian06AA}.

Exact knowledge of the $\gamma$-ray source spectrum resulting from hadronic interactions is thus
of prime importance for a proper physical interpretation. The production of neutral pions 
and their subsequent decay to $\gamma$-rays is thought to be the 
principal mechanism for the $\gamma$-ray hadronic component in cosmic ray
interactions, and a number of parameterizations have been developed over the years
that thus describe $\gamma$-ray production in $pp$-collision \cite{Stephens81,Blattnig00,Huang05ICRC,DomingoSantamaria05,
Kamae05,Kamae06,Kelner06}. In addition to the neutral pion production, however, the direct 
$\gamma$-ray production in cosmic-ray interaction was found important \cite{Huang05ICRC}, as was
the production of $(\Sigma^\pm,~\Sigma^0)$, $(K^\pm,~K^0)$, and $\eta$ particles, all of which eventually decay 
into $\gamma$-rays \cite{Aharonian00,Gaisser90}. Also, the helium nuclei in cosmic rays and the interstellar medium
are expected to contribute about 20-30\% to the secondary production in cosmic-ray interactions, for which the
multiplicity spectra may be different.

The purpose of this work is to carefully study the $\gamma$-ray production from individual decay processes, 
including the direct production, in cosmic-ray interactions. For that purpose we employ the 
event generator DPMJET-III \cite{Roesler00} to simulate secondary productions in both p-generated and 
He-generated interactions. We include all relevant secondary particles with $\gamma$-rays as the 
final decay products. For the cosmic rays, protons and helium nuclei are taken into account. 
The composition the ISM we assume as 90\% protons, 10\% helium nuclei, 0.02\% carbon, and 0.04\% oxygen. 
At energies below 20 GeV, where DPMJET appears unreliable, we combine the simulation results with 
a parametric model of $\pi^0$ production \cite{Kamae06}, that includes the resonances 
$\Delta$(1232) and $\Delta$(1600), thus deriving a $\gamma$-ray production matrix for cosmic rays with energies 
up to about 10 PeV that can be easily used to interpret
the spectra of cosmic $\gamma$-ray sources.  

As examples we use the $\gamma$-ray production matrix to interpret the observed
spectra of diffuse galactic emission and of one particular supernova remnant.
We calculate the spectrum of diffuse galactic $\gamma$-rays and find, as other groups did before
\cite{Kamae05}, that the GeV excess is probably not the result of an inappropriate model of hadronic
$\gamma$-ray production. We also test the hypothesis that the TeV-band $\gamma$-ray emission of the 
shell-type SNR RX J1713-3946, that was observed with HESS, is caused by hadronic 
cosmic rays that have a spectrum according to current theories of cosmic-ray modified shock acceleration.
\section{Cosmic-Ray Generated Hadronic Interactions}
\subsection{Parameterized $\pi^0$ production in $pp$ collisions}
The $\pi^0$ production and subsequent decays are thought to be one of the most principal $\gamma$-ray 
production mechanisms in high-energy astrophysics. The $\pi^0$
production in $pp$-collisions was well studied both theoretically and experimentally. However, all 
the current available models have met an applicability limit: 
the isospin model \cite{Lipkin72}, considering the relation 
$\frac{d\sigma_{\pi^0}}{dp}=\frac{1}{2}\left(\frac{d\sigma_{\pi^+}}{dp}+\frac{d\sigma_{\pi^-}}{dp}\right)$,
is found to be too simplified by comparison with accelerator data \cite{Lipkin72,Jaeger75I,Jaeger75II,Buesser76} 
for incident proton energies from tens of GeV to 
a few hundred of GeV, and the parameterized $\pi^+$ and $\pi^-$ production cross sections in proton-induced 
nuclear collisions are found to deviate from the experimental data at the incident proton energy 
$E_p>200$ GeV or at the transverse momentum $p_T>2$ GeV/c \cite{Blattnig00,Liu03}; 
the scaling model prediction \cite{Stephens81} is found to give a poor fit in both differential 
and total cross sections at low energy $E_p\le 10 $ GeV \cite{Dermer86AA},
and it was also shown \cite{Tan83} that the kinematic limit of this model is incorrect at energy $E_p \le 3$ GeV; 
the isobar model \cite{Dermer86APJ,Tan83,Stecker70} is found superior
but is valid only from threshold up to a few GeV \cite{Dermer86APJ}. 

In addition to the theoretical approach, the parametrization of the inclusive cross section of $\pi^0$ 
production in $pp$-collisions based on the accelerator data was also 
studied \cite{Stephens81,Blattnig00,Kamae06}. The validity of a parametrization depends on the 
functional formulae chosen and also the data used to
determine the free parameters. Thus, the parametrization method often re-produces tails extending 
beyond the energy-momentum conservation limit, 
which might result in unphysical structures when the spectrum is calculated in a broad energy range. 
It is recognized \cite{Huang05ICRC,DomingoSantamaria05} that these parameterizations lose accuracy in the
high energy regime and come with a kinematic limit of validity. The parametrization of Stephens and Badhwar 
\cite{Stephens81} under-predicts the yield of high energy pions. 
The work of Blattnig {\it et al.} \cite{Blattnig00} is valid only for energies $E_p$ below 50 GeV, 
and above this energy it extremely over-predicts the pion production 
and carries a kinematic invalidity; Kamae {\it et al.} \cite{Kamae06} have exercised much care in 
determining the inclusive cross section near the pion production threshold, 
but the same problems with kinematic invalidities, etc, arise in the high energy range.

The spectral distribution of secondary particles and the total cross section are not directly 
obtainable due to the lack of data. Nevertheless, one can calculate the energy spectrum or total 
inclusive cross section by integrating the parameterized differential inclusive cross section 
over the whole phase space (the scattering angle and the energy). By analyzing the behavior of the
integral, it is possible to evaluate the accuracy and validity limit of different parameterizations. 
For example, by the analysis of the total cross section, a parametrization which cannot integrate 
to a correct total cross section can be ruled out, even though the invariant differential cross 
section data can be reproduced, since the global behavior of the parametrization is not accurate. 
On the other hand, producing a correct total cross section upon integration doesn't necessarily 
imply accuracy of the global behavior of this parametrization. 
This disadvantage results from the fact that the available data are not sufficiently constraining 
the parametrization of differential cross sections.

\subsection{Cosmic-ray generated hadronic interactions by Monte Carlo event generator DPMJET-III}\label{Sec:HadronicReaction}
In particle physics, the Monte Carlo simulation approach is widely applied to obtain information
on particle production in hadronic interactions. The principle advantage of using the Monte Carlo method is that 
the codes can be used directly for calculations of all secondary products. However, it
is quite useful to have simple representations to investigate the fundamental spectral characteristics.
In this work, the Monte Carlo event generator, DPMJET-III, is used to simulate 
the secondary production in cosmic-ray generated hadronic interactions.

The Monte Carlo event generator DPMJET \cite{Roesler00}, is constructed for sampling hadron-hadron, 
hadron-nucleus, nucleus-nucleus and neutrino-nucleus interactions from a few GeV up to very high
energies. DPMJET is an implementation of the two-component Dual Parton Model for the description of interactions 
involving nuclei, including photonproduction and deep inelastic scattering off nuclei. It is based 
on the Gribov-Glauber approach and treats both soft and hard scattering processes in an unified way. 
Soft processes are parameterized according to the Regge-phenomenology whereas lowest-order perturbative 
QCD is used to simulate the hard component. Multiple-parton interactions in each individual 
hadron/nucleon/photon-nucleon interaction are described by the PHOJET event generator. 
The fragmentation of parton configurations is treated by the Lund model PYTHIA.

As a new feature, DPMJET-III allows the simulation of enhanced graph cuts in non-diffractive inelastic 
hadron-nucleus and nucleus-nucleus interactions. For example, in an event with two wounded nucleons, 
the first nucleon might take part in a non-diffractive interaction whereas the second one 
scatters diffractively producing only very few secondaries. Further features of DPMJET-III are 
a formation-zone intranuclear cascade and the implementation of certain baryon stoppings.

\begin{figure}
\mbox{\includegraphics[scale=0.45]{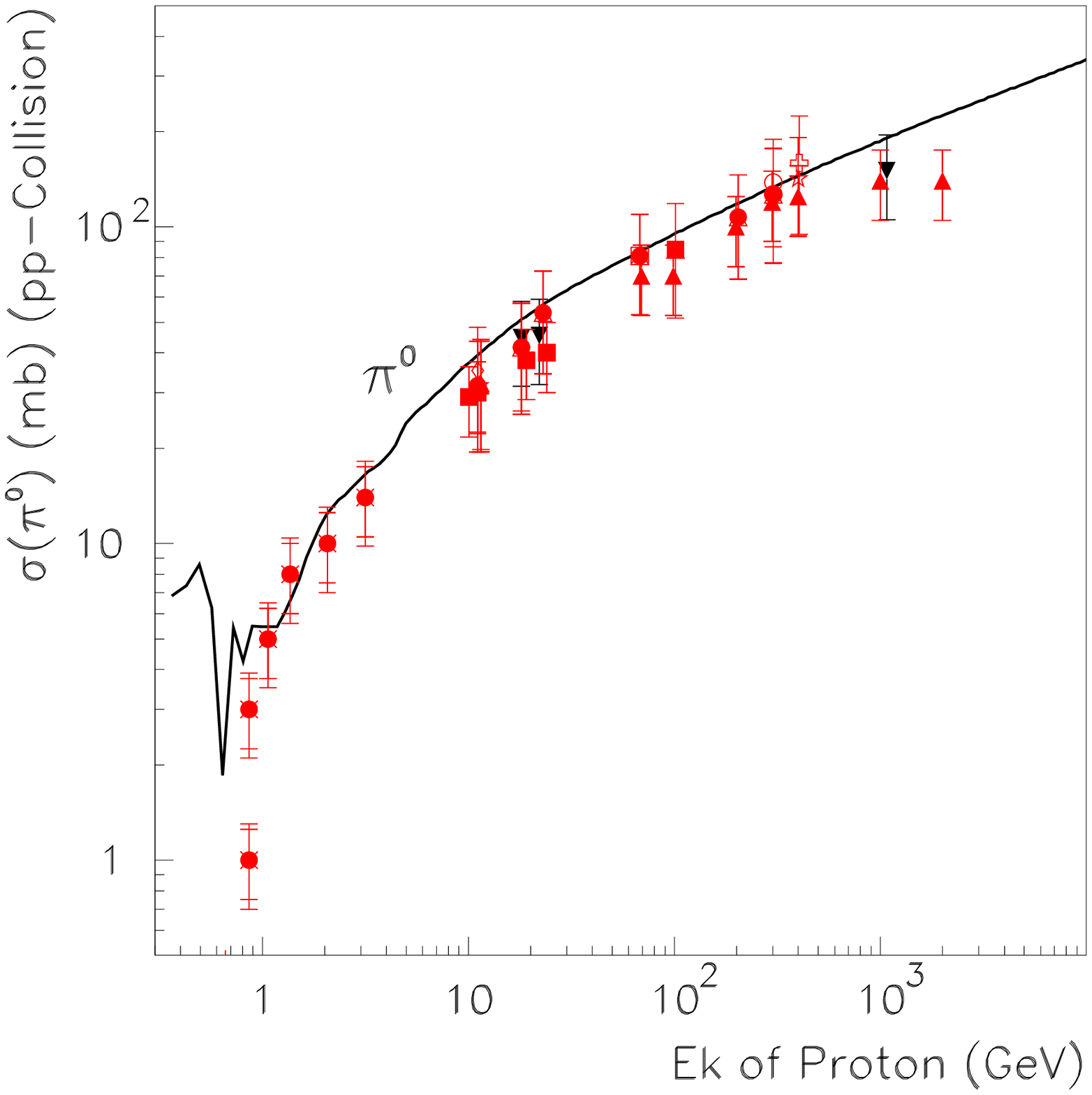}
      \includegraphics[scale=0.45]{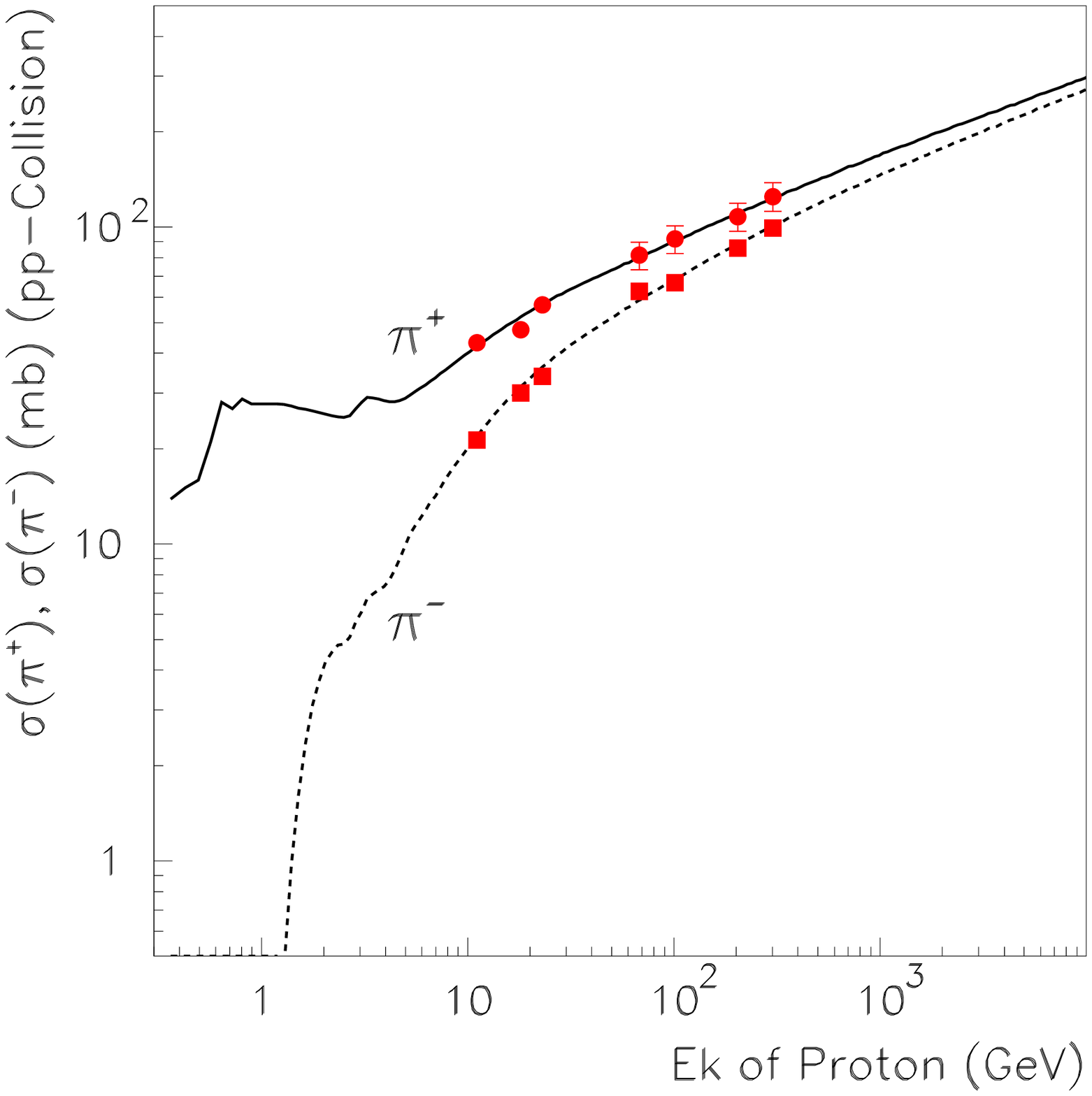}}
\caption{$\pi^0$ and $\pi^\pm$ production cross sections in pp collisions simulated by DPMJET-III,
here shown as function of the kinetic energy $E_k$ of incident 
protons. 
The data shown figure are quoted from \cite{Stephens81,Jaeger75I,Kass79}}\label{Fig:DPMJETXSection}
\end{figure}

DPMJET-III and earlier versions such as DPMJET-II have been extensively tested against data in 
hadron--nucleus and nucleus--nucleus collisions \cite{Roesler00,Ranft95}, for the 
simulation of cosmic-ray air shower problems \cite{Knapp03}, and for the atmospheric $\gamma$-ray and neutrino
production \cite{Kasahara02}. Although DPMJET has reproduced well the experimental data in hadronic collisions 
over a quite broad incident energy range, it appears to produce unphysical structure near the threshold
energy of pion production. Figure~\ref{Fig:DPMJETXSection} shows the $\pi^0$ and $\pi^\pm$ production cross 
sections in function of the kinetic energy of protons in $pp$-collisions
simulated by DPMJET-III. The experimental values are taken from references \cite{Stephens81,Jaeger75I,Kass79} 
up to $E_p=2~\textrm{TeV}$. To be noted from the figure is that
the DPMJET results agree with the data except the energy range $E_k \le 2\sim 3~\textrm{GeV}$. 
For a high-energy physics event generator this is not uncommon behavior. The energy region 
in question corresponds to the resonance region relating to the nuclear cascades in the nuclear interactions. 
The contribution from the baryon resonances, either $\Delta(1232)$ or $\Delta(1600)$, has a very 
narrow spike-like characteristic at energies close to pion production threshold, and is sometimes added by hand
to parameterizations of the differential cross section \cite{Kamae06}.

We therefore use the parameterizations \cite{Blattnig00,Kamae06} including pion production and resonance 
production below a total energy 
$E_p < 20~\textrm{GeV}$. For $E_p > 20~\textrm{GeV}$, the production is given by the DPMJET-III simulation; 
for $E_p < 2.5~\textrm{GeV}$, it is purely parametric with
resonance production included; for $2.5~\textrm{GeV} \le E_p \le 20~\textrm{GeV}$, a linear combination in 
energy is used to transition from the parametrization to the 
DPMJET-III results. The final production spectra of secondary particles are smooth and still retain agreement 
with experimental data.

We use DPMJET to calculate energy-dependent weight factors, that allow us to
parametrically account for $p+ISM$ and $He+ISM$ collisions in the parametrization approach, that has been 
strictly derived for $p+p$ collisions only.
The weights show no strong dependence on energy of the projectile particle.

A remark is in order drawn on the parametrization of pion production, that we use at low cosmic-ray energies. The functional 
form of the pion multiplicity spectra violates kinematic limits, so typically a few percent of
pions appear to have energies higher than possible. A truncation of the multiplicity spectra at the kinematic limit
would modify the mean energy of pions, thus changing the $\gamma$-ray spectra.
As the parametrization is a best fit to data, any modification would in principle deteriorate the fit.
We therefore decided not to correct for the violation of kinematic limits.
\subsection{Decay channels of secondary particles to $\gamma$-rays}\label{Section:DecayMode}

The isobar model is based on the theory that $\pi$ production in $p+p$ collisions near the threshold is mediated 
by the excitation of the $\Delta_{3/2}$ isobar, which
subsequently decays into a nucleon and pions. In this work, the produced isobar is assumed to 
carry momentum directly forward and backward in the
center-of-mass system of the collisions with an equal probability, and then decays isotropically into a nucleon and $\pi^0$'s.

The decay modes and decay fractions for resonances are considered as follows based on the Particle Data Group.
\begin{itemize}
\item[] $\Delta(1232)       \rightarrow   p                +      \pi^0$;
\item[] $\Delta(1600)       \rightarrow   p                +      \pi^0$;
\item[] $\Delta(1600)       \rightarrow   \Delta(1232)     +      \pi^0$;
\item[] $\Delta(1600)       \rightarrow   \textrm{N(1440)} +      \pi^0$;
\item[] $\textrm{N(1440)}   \rightarrow   p                +      \pi^0$;
\item[] $\textrm{N(1440)}   \rightarrow   \Delta(1232)     +      \pi^0$;
\item[] $\textrm{N(1440)}   \rightarrow   p                +      2\pi^0$.
\end{itemize}
The $\pi^0$ contribution from resonances is then added to the non-resonant $\pi^0$ production.

While running the DPMJET-III, all secondary particles produced in simulated hadronic interactions are recorded. 
The following decay modes shows all the decay processes of secondary products with $\gamma$-rays 
as the final decay particles that are taken into account in the current work.
\begin{itemize}
\item baryonic decays:
\begin{itemize}
\item[] $\Lambda            \rightarrow   n             +      \pi^0$;
\item[] $\bar{\Lambda}      \rightarrow   \bar{n}       +      \pi^0$;
\item[] $\Sigma^0           \rightarrow   \Lambda       +      \gamma$;
\item[] $\Sigma^+           \rightarrow   p             +      \pi^0$.
\end{itemize}
\end{itemize}
\begin{itemize}
\item mesonic decays:
\begin{itemize}
\item[] $\pi^0              \rightarrow   2 \gamma$;
\item[] $\pi^0              \rightarrow   e^-           +      e^+    +      \gamma$;
\item[] $K^+                \rightarrow   \pi^+         +      \pi^0$;
\item[] $K^-                \rightarrow   \pi^-         +      \pi^0$;
\item[] $K^0_S              \rightarrow   2 \gamma$;
\item[] $K^0_L              \rightarrow   3 \pi^0$;
\item[] $K^0_L              \rightarrow   \pi^+         +      \pi^-  +      \pi^0$.
\end{itemize}
\end{itemize}
For the two-body decay process, it is easy to evaluate by particle kinematics while for a 
three-body decay process, 
the Dalitz-plot is used. 
For the fraction of each individual decay process, the values published in Particle Data Group are applied. 

\begin{figure}
\mbox{\includegraphics[scale=0.45]{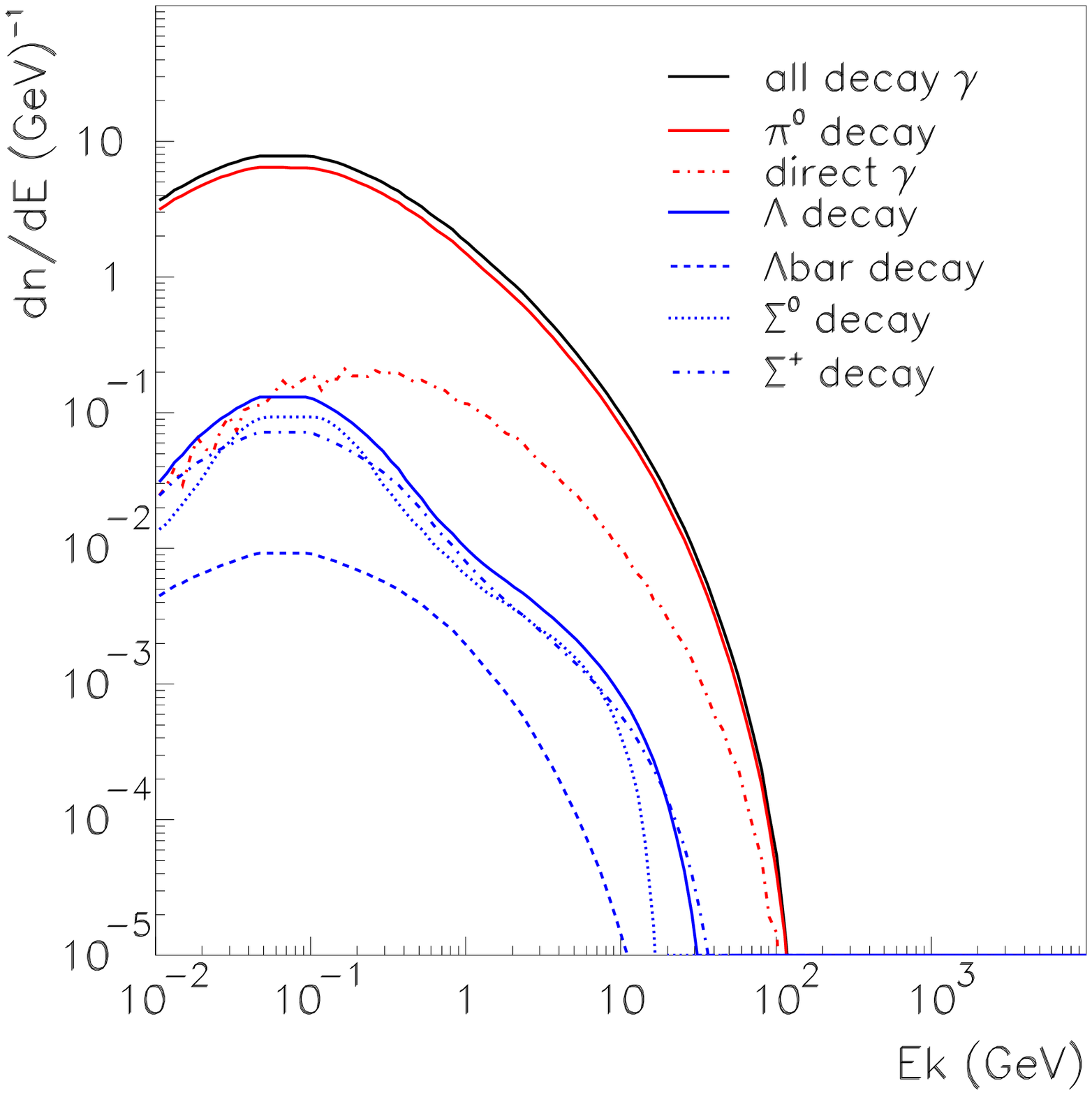}
      \includegraphics[scale=0.45]{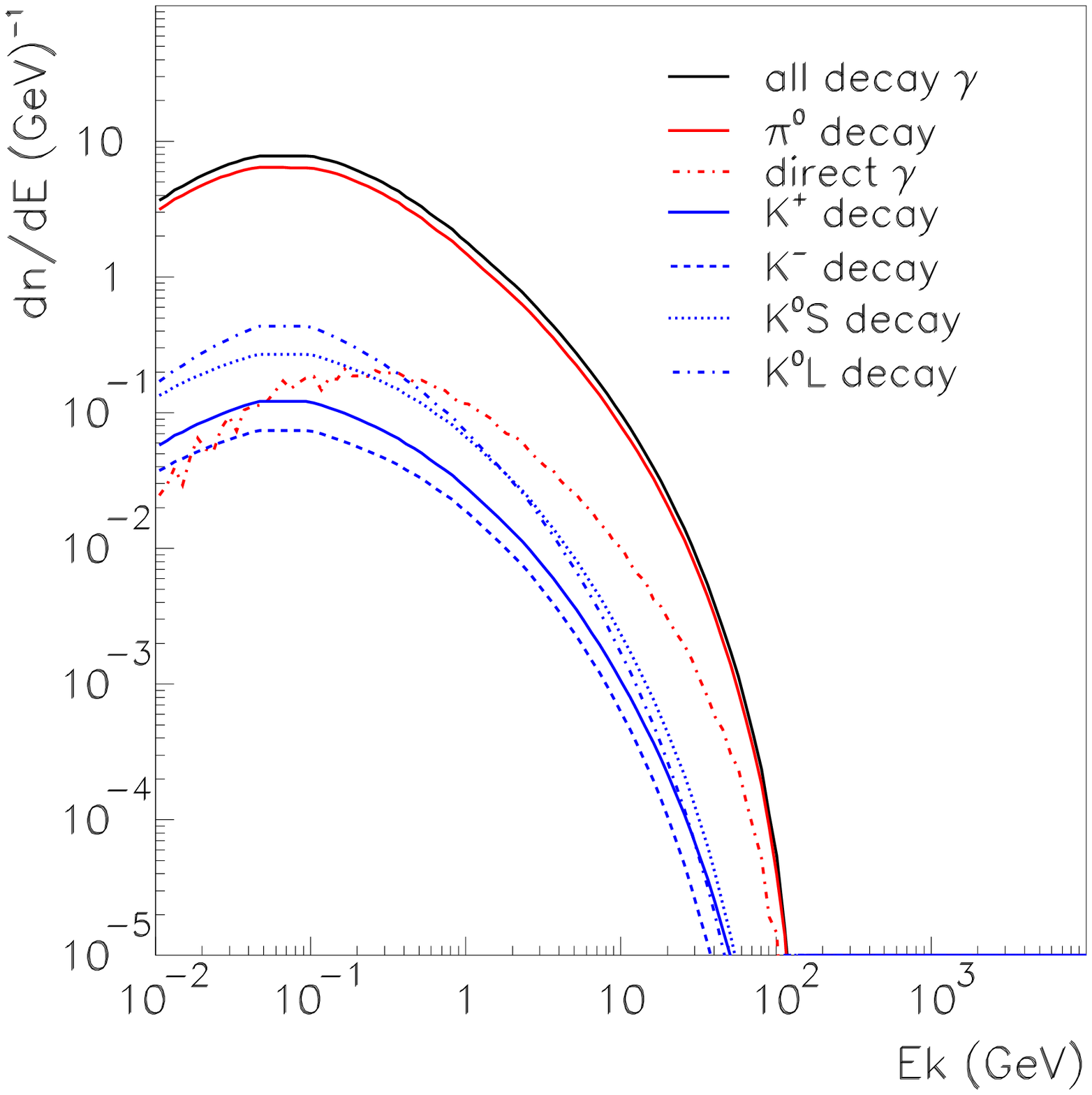}}\\
\mbox{\includegraphics[scale=0.45]{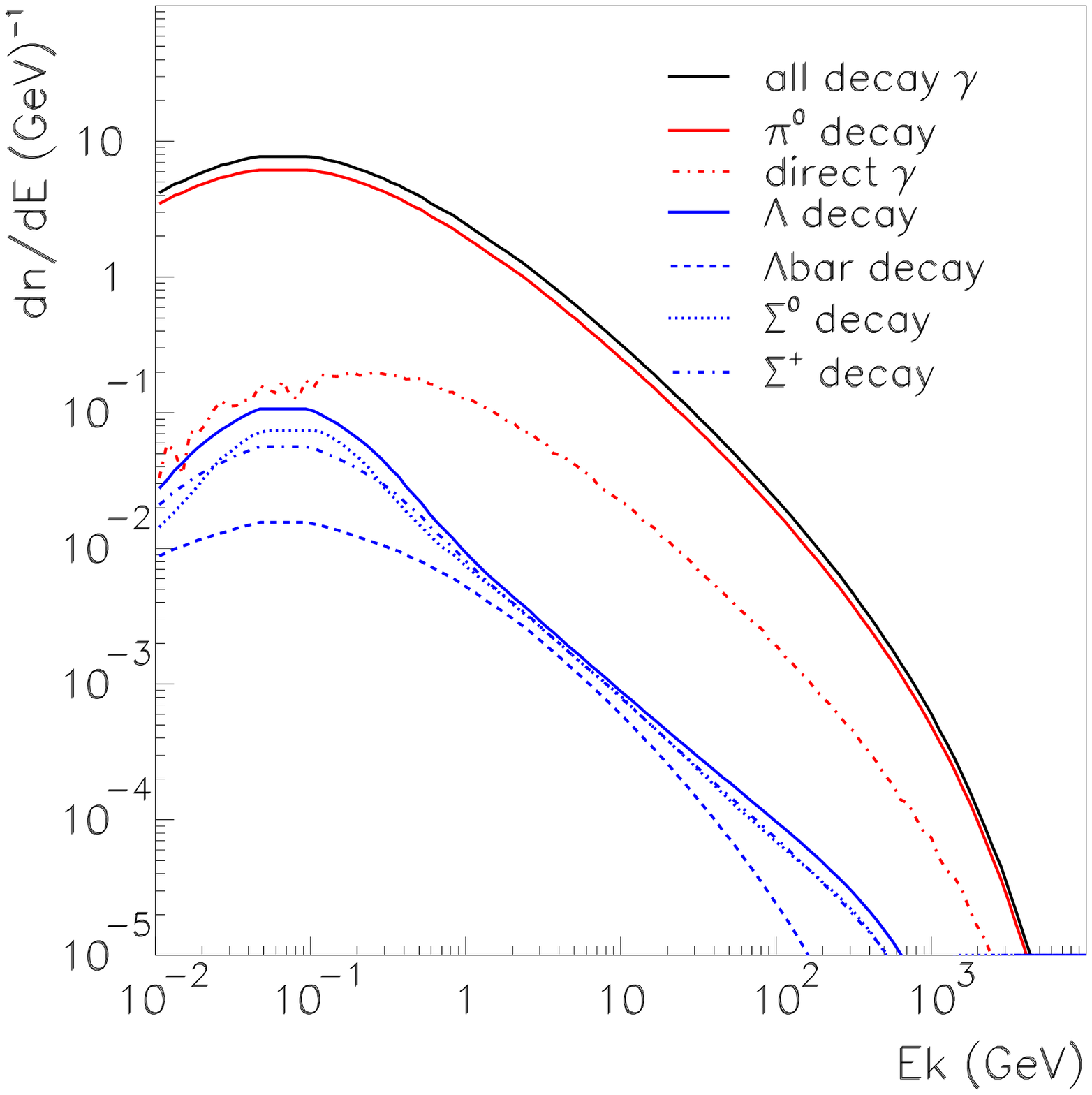}
      \includegraphics[scale=0.45]{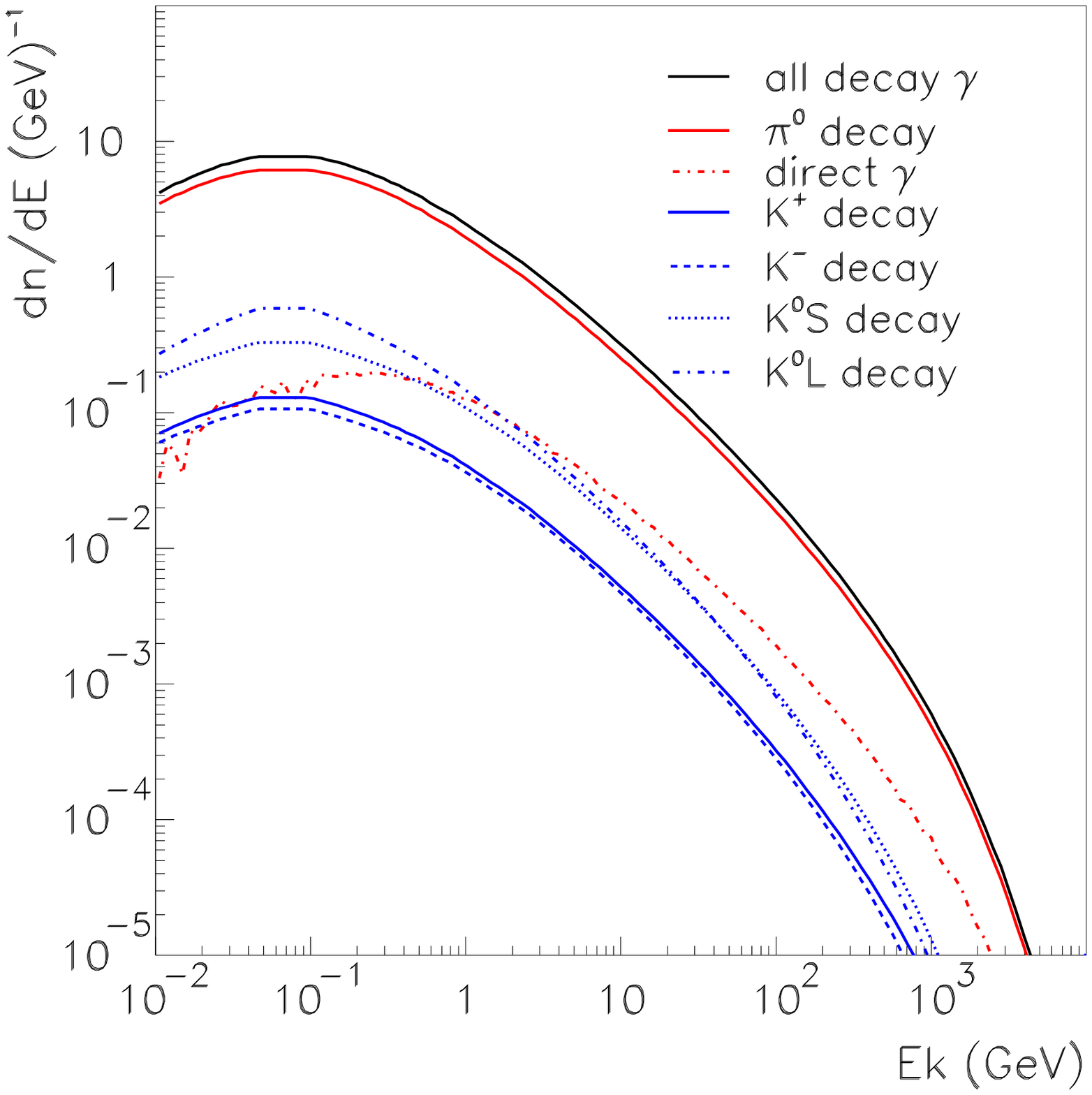}}
\caption{The energy spectra of the $\gamma$-ray contributions from each secondary particle produced in $p+ISM$ 
interactions for two different incident proton energies: 150 GeV (upper panel) and 10 TeV (lower panel). For ease 
of comparison, baryonic (left) and mesonic (right) decay contributions are separated. In each figure, 
the energy spectrum of total $\gamma$-rays (the sum of all contributions) and also the spectrum of the 
directly produced $\gamma$-ray photons are shown.}\label{Fig:DecaySpectra}
\end{figure}

Figure~\ref{Fig:DecaySpectra} shows the energy spectra of $\gamma$-rays contributed from each type of 
secondary particles produced in p+ISM interactions for two different 
incident proton energies: 150 GeV (upper curve) and 10 TeV (lower curve). The ISM composition is taken 
as 90\% protons, 10\% helium nuclei, 0.02\% carbon, and 0.04\% oxygens. 
For ease of comparison, baryonic (left) and mesonic (right) decay contributions are separately shown. 
The total energy spectra of $\gamma$-rays 
(the sum of each contribution), the directly produced $\gamma$-ray photons in the hadronic collisions and 
the $\gamma$-rays from $\pi^0$ decays are also plotted in each 
figure for comparison. To be noted in the figures is that, in addition to $\pi^0$ decays, the 
directly produced $\gamma$-ray photons appear more important than other mesons and baryons over almost 
the whole energy domain. However, the secondary mesons contribute more at the peak of the $\gamma$-ray spectrum. 
Integrated over the $\gamma$-ray spectrum, the other secondary particles have contributed about 
20\% $\gamma$-ray photons relative to the $\pi^0$ decays.

Some authors \cite{Gaisser90,Kelner06} suggested that a few percent of $\gamma$-rays should be
contributed by decays of $\eta$ mesons, but no $\eta$ meson is 
listed in DPMJET read-out table. This is due to the fact that with our
simulation setup, the $\eta$ mesons have already decayed and only their decay products are listed.
We have verified that the $\eta$ mesons are properly accounted for by performing test runs
under different PYTHIA parameters, so the $\eta$-decay is avoided. Another independent verification 
arises from the very small contribution to $e^-/e^+$ and $\mu^-/\mu^+$ pairs from $\eta$ decays, which are found 
consistent within the statistical uncertainty with all $\eta$-mesons having decayed.
\section{Spectrum of $\gamma$-rays Generated by Cosmic Rays}
\subsection{Energy loss in production collisions}
\begin{center}
\begin{figure}
\includegraphics[scale=0.55]{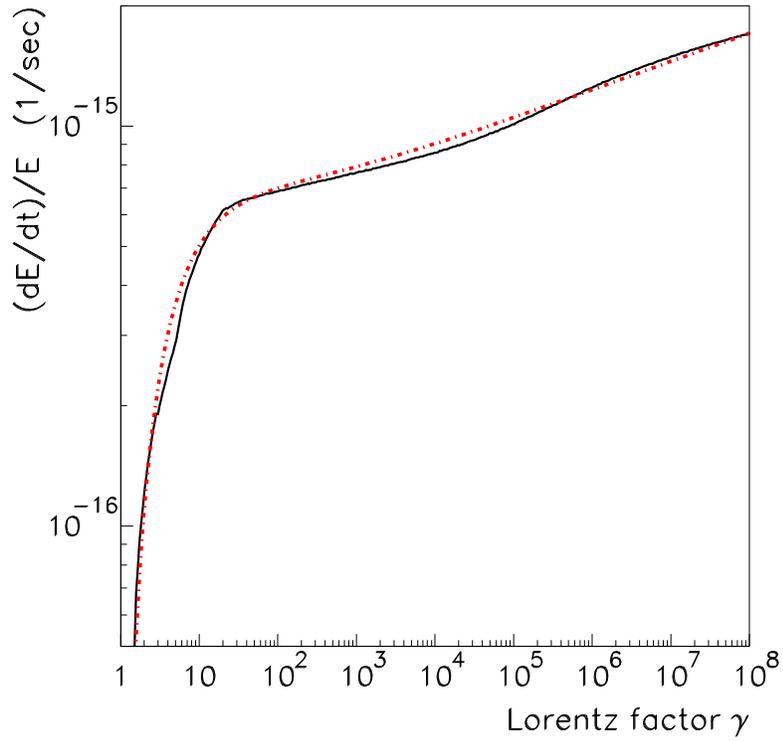}
\caption{The energy loss rate for $p+ISM$ collisions as calculated by integrating the secondary particle yield. 
The thick dash-dot line shows an analytical approximation. The ISM number density is assumed $n_{ISM}=1.0~\textrm{/cm}^3$.}\label{Fig:EDotE}
\end{figure}
\end{center}

The energy loss resulting from the inelastic production collisions can be determined by integrating 
over the yield in all secondary particles. Figure~\ref{Fig:EDotE} 
shows the energy loss rate as function of the incident particle energy in the $p+ISM$ collisions. 
To be noted from the figure is that the energy loss rate gradually increases with energy above 10 GeV on account 
of an increasing importance of
secondary particles other than pions. 
The thick dashed curve in the figure is a functional approximation to the energy loss rate in $p+ISM$ 
production collisions over the 
entire energy range described as 
\begin{eqnarray}\label{EQ:EDot}
\frac{\dot{E}}{E} \simeq  6.78 \times 10^{-16} \frac{(\gamma-1)^2}{\gamma (\gamma +1)}
            - 1.55 \times 10^{-17} \log(\gamma)
                   + 1.80 \times 10^{-17} \log^2 (\gamma).
\end{eqnarray}
To be noted is that the first term in Eq.~(\ref{EQ:EDot}) is a functional approximation 
introduced earlier for pion production 
up to 100 GeV \cite{Pohl00}. This term is a good approximation below 100 GeV where pions are the 
dominant secondary product.
\subsection{The $\gamma$-ray production matrix}
For the cosmic-ray hadronic interactions and the subsequent decays, the spectrum of a final secondary particle 
is described by
\begin{eqnarray}\label{Eq:2ndSpectra}
Q_{2nd}(E) 
= \frac{dn}{dt\cdot dE \cdot dV}
=n_{ISM} \int_{E_{CR}} dE_{CR}\, N_{CR}(E_{CR})\,c\beta_{CR}\left(\sigma \frac{dn}{dE}\right)
\end{eqnarray}
with $N_{CR}(E_{CR}) =\frac{dn_{CR}}{dE_{CR}\cdot dV} ~\textrm{(GeV~cm}^3)^{-1}$ defined as the 
differential density of CR particles ($p$ or $\alpha$). The differential cross section of a secondary 
particle produced in the $(p,\alpha)+ISM$ collision is 
\begin{eqnarray}\label{Eq:ProdXSection}
\frac{d\sigma}{dE}(E_{CR},E) = \sigma_{prod} \frac{dn}{dE} 
\end{eqnarray}
where $\sigma_{prod}$ is the inelastic production cross section, whose value is simulated by DPMJET,
and $\frac{dn}{dE}$ is the multiplicity spectrum of the secondary particle in question.

We can follow the decay of unstable secondary particles into $\gamma$-rays, which is a purely kinematical problem, 
and thus obtain the $\gamma$-ray spectrum from secondary-particle decay generated by a 
cosmic-ray particle with energy $E_{CR}$ as
\begin{equation}
Q_{\gamma}(E_\gamma) 
= \sum_k n_{ISM}\int_{E_{CR}} dE_{CR}\ N_{CR}(E_{CR})\, c\beta_{CR}\, \sigma (E_{CR})\,
\frac{dn_{k,\gamma}}{dE_\gamma}(E_\gamma, E_{CR})
\label{Eq:gaSpectra}
\end{equation}
where $\frac{dn_{k,\gamma}}{dE_\gamma}$ is the $\gamma$-ray spectrum resulting from the initial multiplicity
spectrum of the secondary particle species $k$ or the directly produced $\gamma$-ray spectrum in cosmic-ray
generated interactions. 

If the total energy of cosmic-ray particles ($p$ or $\alpha$) is parameterized as
\begin{eqnarray}\label{Eq:ECR}
E_T = 1.24\cdot (1+0.05)^j \quad \textrm{GeV/n}
\label{CRenergy}
\end{eqnarray}
and the $\gamma$-ray energy is sampled as
\begin{equation}
E_\gamma =0.01\cdot (1.121376)^{i-0.5} \quad \text{GeV} ,
\label{gaenergy}
\end{equation}
both with sufficient energy resolution, then the production integral, Eq.~(\ref{Eq:gaSpectra}), can be written as a sum.
\begin{eqnarray}
Q_{\gamma}(E_i) 
&=& \sum_{j} n_{ISM}\, \Delta E_{j}\ N_{CR}(E_j)\,c\beta_j\,\sigma (E_{j})\,\sum_{k}\,\frac{dn_{k,\gamma}}{dE_\gamma}(E_i,E_j)\\
&=& \sum_j n_{ISM} \,\Delta E_{j}\, N_{CR}(E_j)\, c\beta_{j}\, \sigma_j \, \mathbb{M}_{ij}  .\label{Eq:ProductionMatrix}
\end{eqnarray}
The problem is thus reduced to a matrix operation,
in which a vector, that describes the cosmic-ray flux at various energies according to Eq.~(\ref{CRenergy}), is transformed
into a vector composed of the $\gamma$-ray source function at a number of energies {predefined in Eq.~(\ref{gaenergy})}.  

Eq.~(\ref{Eq:ProductionMatrix}) is the final formula with which one can evaluate the $\gamma$-ray spectrum generated 
by cosmic rays. It includes $\sigma_j$, the production cross section at the cosmic-ray ($p$ or $\alpha$) energy $E_j$ 
defined in Eq.~(\ref{Eq:ECR}), and the $\gamma$-ray energy spectrum matrix $\mathbb{M}_{ij}$. Each element of the
matrix, $\mathbb{M}_{ij}$, is the value of the resultant particle energy spectrum 
$\frac{dn}{dE}|_{E_{\gamma}=E_i,E_{CR}=E_j}$, 
with $j$ being the index for the generating cosmic-ray particle ($p$ or $\alpha$) and $i$ being the index indicating the 
$\gamma$-ray energy.

In addition to the $\gamma$-ray production matrix, we have also produced the production matrices of 
all particles that are stable on a time scale relevant for
cosmic-ray propagation, namely
$p,~\bar{p},~e^\pm,~\nu_e,~\bar{\nu}_e,~\nu_{\mu},~{\textrm and}~\bar{\nu}_{\mu}$.
\subsection{Examples of $\gamma$-ray spectra}
With this production matrix, we calculate the $\gamma$-ray spectra for two examples 
of cosmic-ray spectra: a simple power-law spectrum and 
a power-law spectrum with an energy cutoff:
\begin{eqnarray}
N_{CR} &\propto& E_{CR}^{-2.75} \label{Eq:CRSpectraPowerLaw} \\
N_{CR} &\propto& E_{CR}^{-2.75} \cdot \exp \left ( \frac{-E_{CR}}{E_{\textrm{cutoff}}} \right),
\quad\textrm{with}~ E_{\textrm{cutoff}}=~ \textrm{100~TeV}\label{Eq:CRSpectraCutoff}
\end{eqnarray}
$E_{CR}$ being the cosmic-ray total energy
and both assuming the cosmic-ray composed of protons and helium nuclei.

\begin{figure}
\mbox{\includegraphics[scale=0.45]{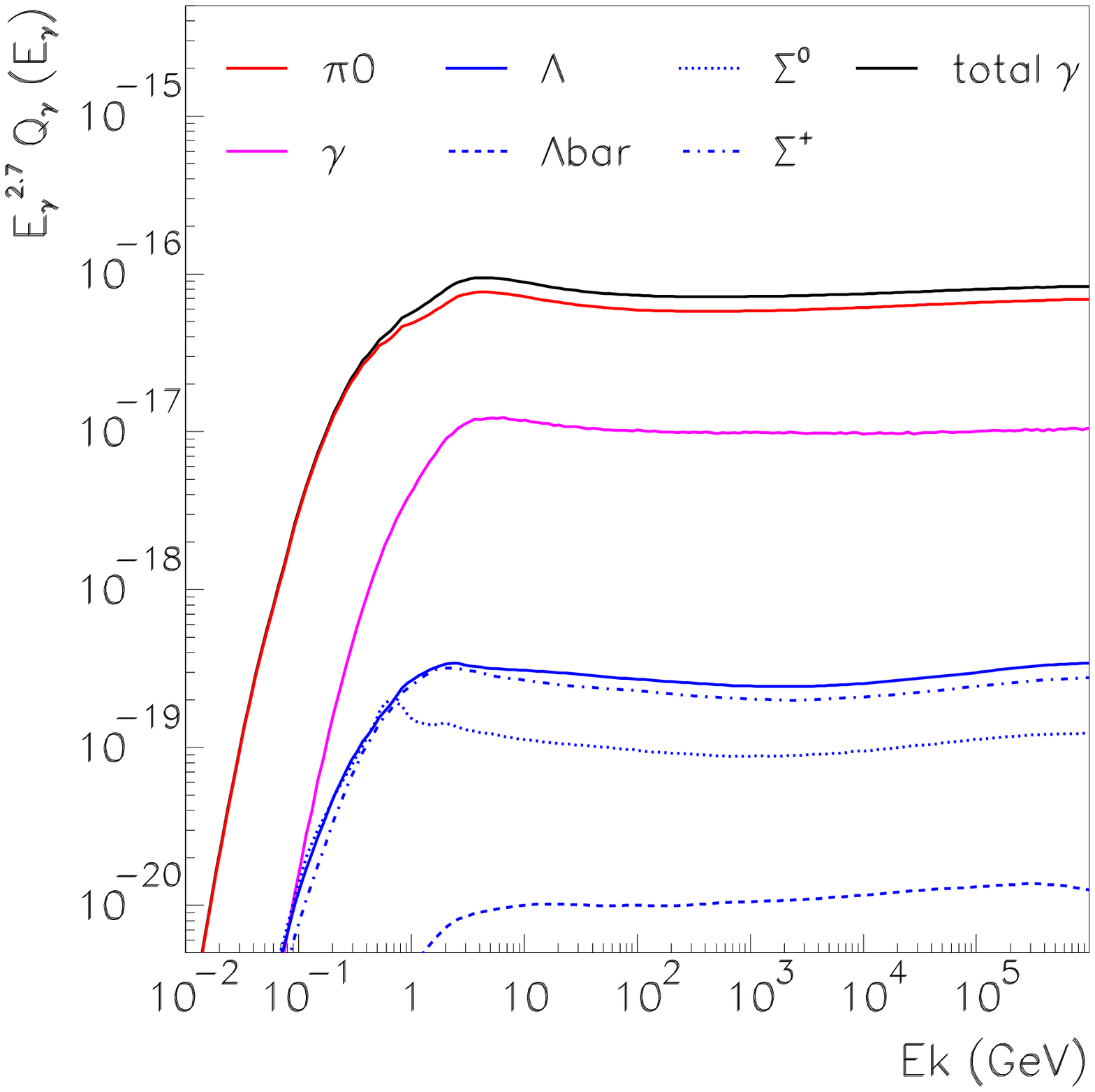}
      \includegraphics[scale=0.45]{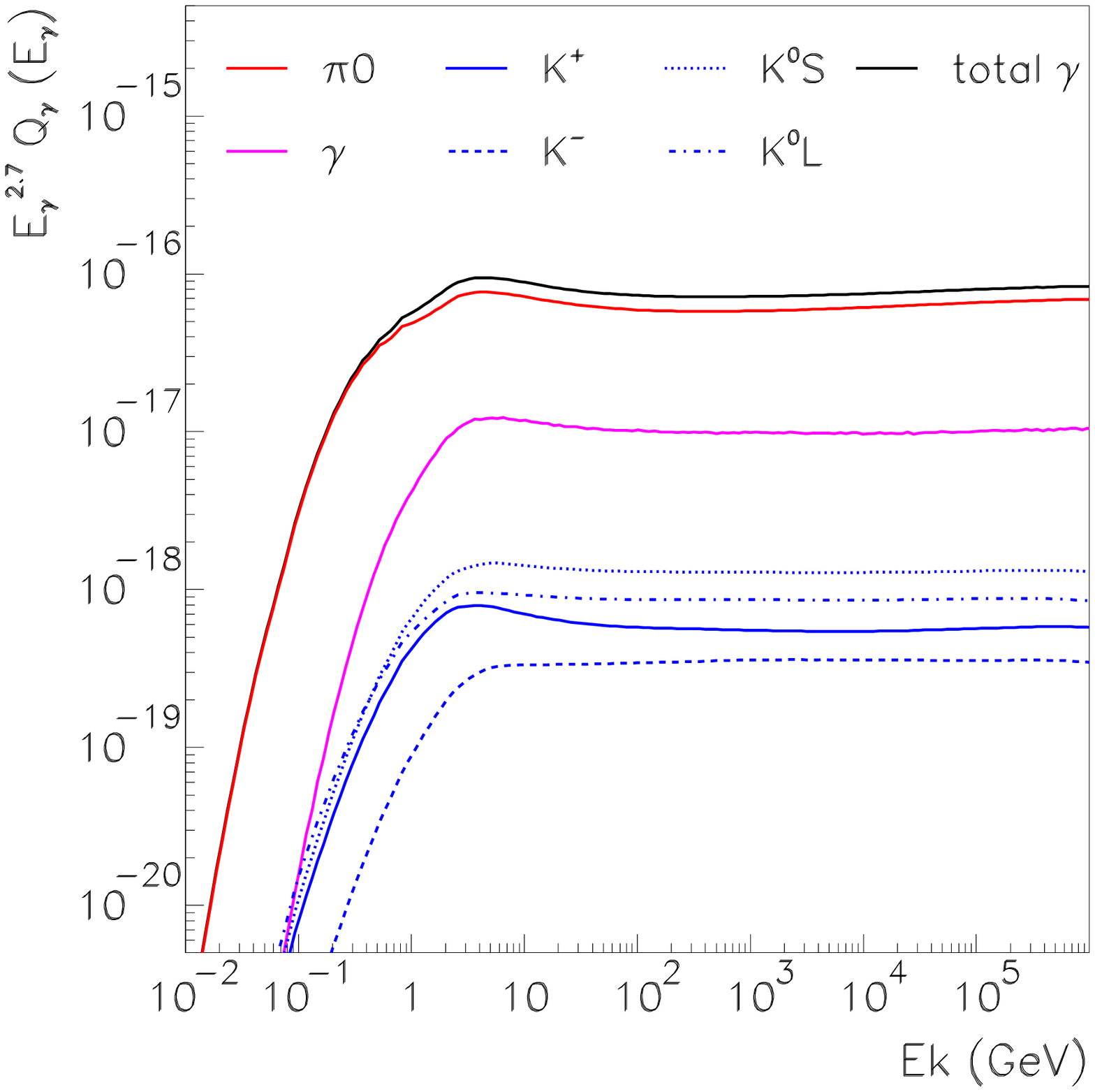}}
\caption{The energy spectra of $\gamma$-rays contributed by the secondaries produced in $CR~(p~\textrm{and}~\alpha)+ISM$ 
interactions with a simple power-law cosmic-ray spectral
distribution in (\ref{Eq:CRSpectraPowerLaw}). Figures are shown for baryonic (left) and mesonic (right) contributions. 
The total $\gamma$-ray spectrum, the components 
from $\pi^0$ decays and the direct $\gamma$ production are also shown in each figure.
Note that we only sample cosmic rays up to $10^8$~GeV.}\label{Fig:GammaSpectraCRPowerLaw}
\end{figure}

\begin{figure}
\mbox{\includegraphics[scale=0.45]{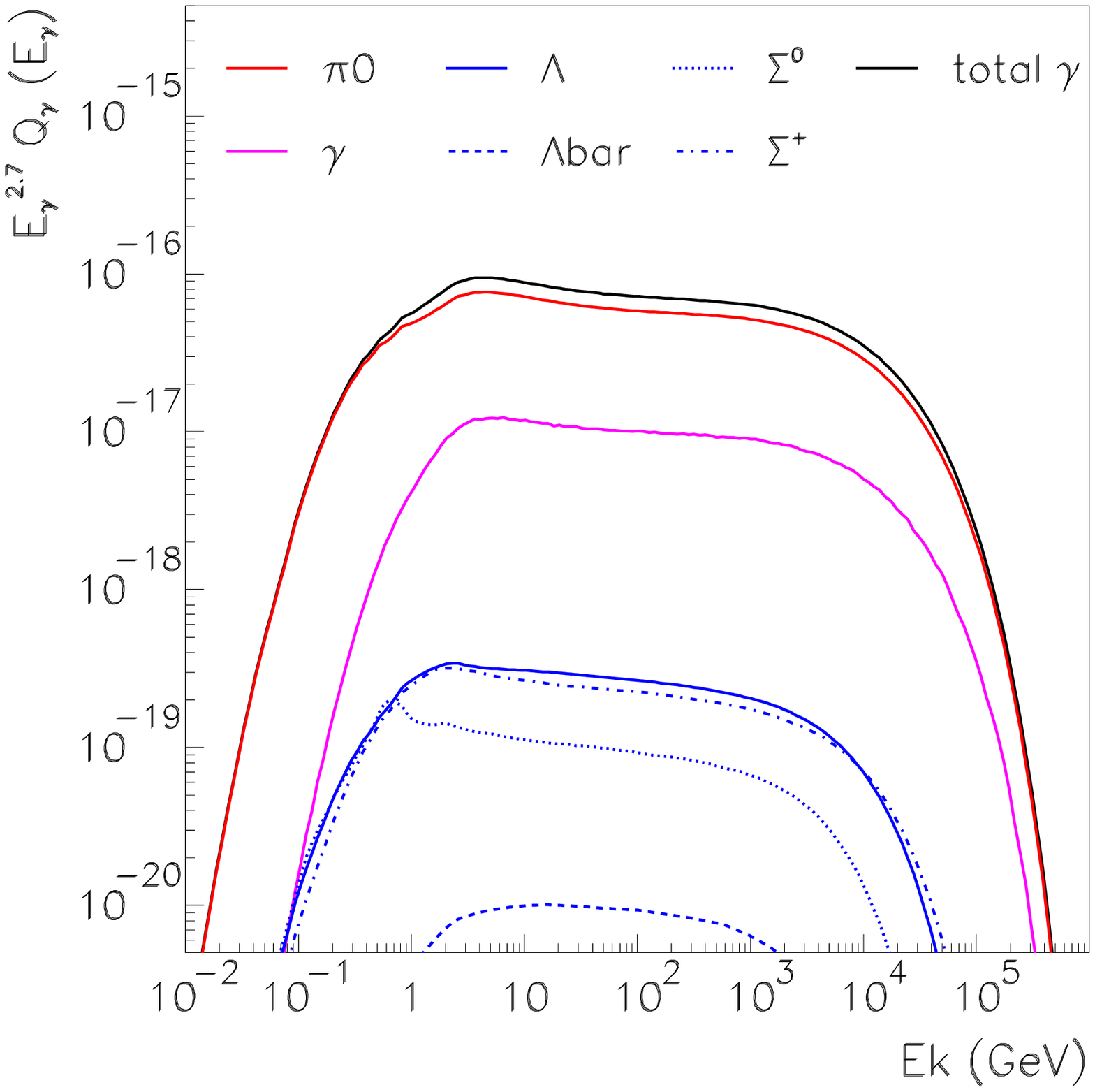}
      \includegraphics[scale=0.45]{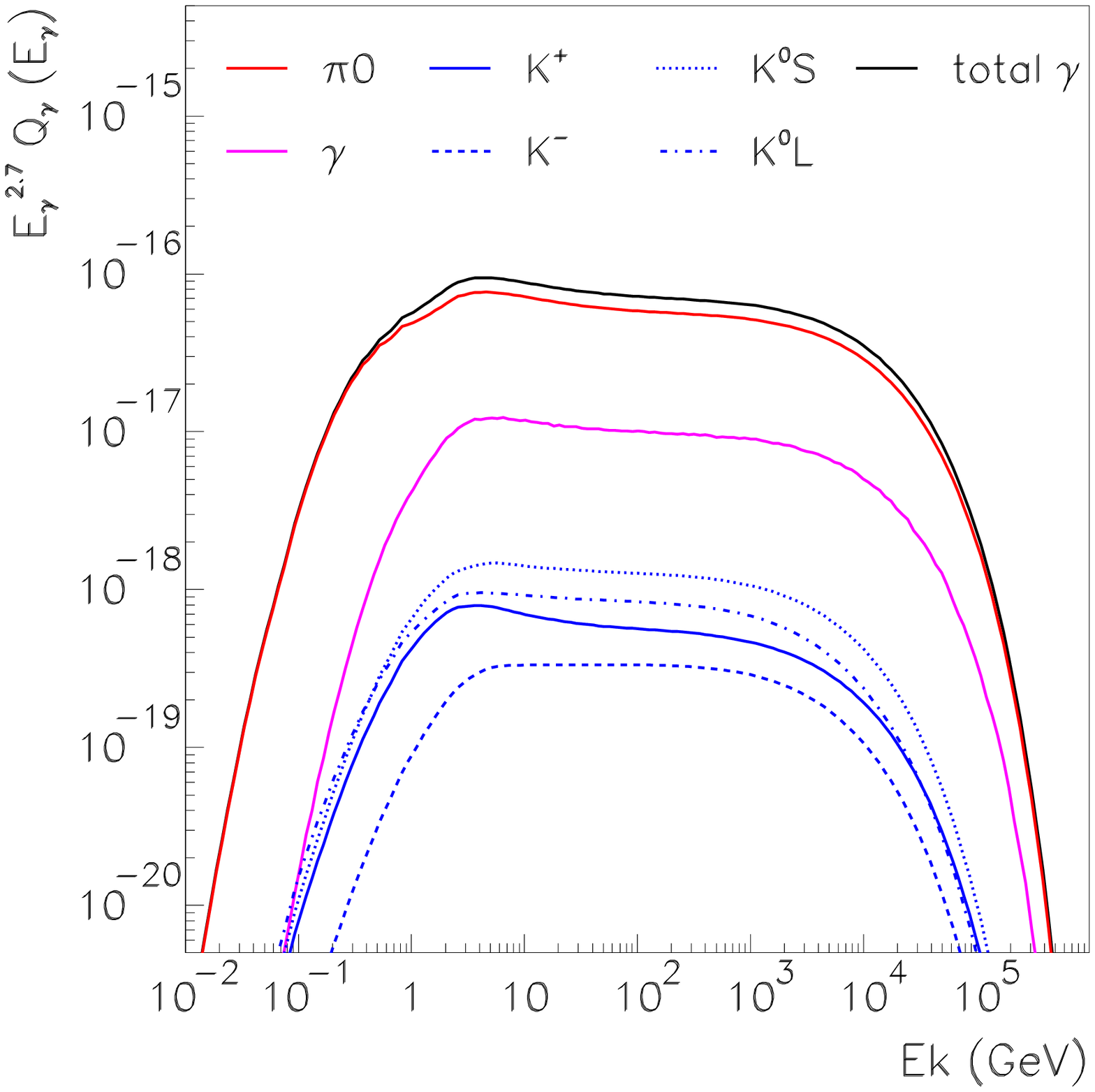}}
\caption{The energy spectra of $\gamma$-rays contributed by secondaries produced in $CR~(p~\textrm{and}~\alpha)+ISM$ 
interactions, assuming an exponential cutoff in the cosmic-ray spectrum at 100~TeV described in Eq.~(\ref{Eq:CRSpectraCutoff}). 
The plots are arranged as in Figure~\ref{Fig:GammaSpectraCRPowerLaw}.}\label{Fig:GammaSpectraCRCutoff}
\end{figure}

Figure~\ref{Fig:GammaSpectraCRPowerLaw} shows the energy spectra of $\gamma$-rays contributed by the secondaries 
produced in $CR~(p~\textrm{and}~\alpha)+ISM$ interactions with 
a simple power-law cosmic-ray spectral distribution described in (\ref{Eq:CRSpectraPowerLaw}). 
Figure~\ref{Fig:GammaSpectraCRCutoff} shows the same
as Figure~\ref{Fig:GammaSpectraCRPowerLaw} but with an energy cutoff in cosmic-ray spectrum at 
$E=~\textrm{100~TeV}$. The contribution from decays of secondary baryons and 
mesons produced in cosmic-ray generated interactions on ISM are separately plotted for comparison. 
In each figure, the total $\gamma$-ray spectrum and the components 
from $\pi^0$ decays and the direct $\gamma$ production are also shown. As shown in the figures, the 
$\pi^0$ decay is the dominant component in the diffuse $\gamma$-rays
spectrum, however, at least about 20\% of the $\gamma$-ray photons result
from secondary products other than pions, 
including the direct $\gamma$-ray production. With the energy cutoff in cosmic-ray spectrum, it is found 
that the $\gamma$-ray spectrum is modified also at energies below 1~TeV, where the spectra in
Figure~\ref{Fig:GammaSpectraCRCutoff} are notably softer than those shown in Fig~\ref{Fig:GammaSpectraCRPowerLaw}.
This is because the multiplicity spectrum of a secondary 
particle has a long tail, and a high energy cosmic-ray particle can 
generate several generations of production collisions, multiplying the production at lower energies. 

\subsection{The $\gamma$-ray spectrum in the GeV band}\label{Section:GeVBump}
The observed GeV-bump in diffuse $\gamma$-ray spectrum \cite{Hunter97} is a long-term standing mystery in 
$\gamma$-ray astrophysics. Both hadronic and leptonic scenarios have been proposed to explain the GeV excess. A harder 
spectrum of cosmic-ray nucleons
is required for $\gamma$-rays of nucleonic origin 
to account for the observed intensity in the multi-GeV range \cite{Aharonian00}, without violation 
of the observed intensity at a few hundred MeV. However, significant variations of the cosmic-ray
spectrum in the Galaxy are unlikely \cite{Buesching05}. Strong fluctuations in the spectrum of cosmic-ray 
electrons must be expected, though \cite{Pohl98,Pohl03}: Electrons are accelerated by SNR with power-law spectra 
of index -2.0 or similar, but severe energy losses prevent them from homogeneously filling the 
Galaxy. A hard spectral component from inverse 
Compton scatterings could then explain the GeV bump in the diffuse $\gamma$-rays. It has also been argued 
that both a harder electron and a harder nucleon spectrum may be required to reproduce the GeV excess
\cite{Strong04}.

\begin{center}
\begin{figure}
\includegraphics[scale=0.6]{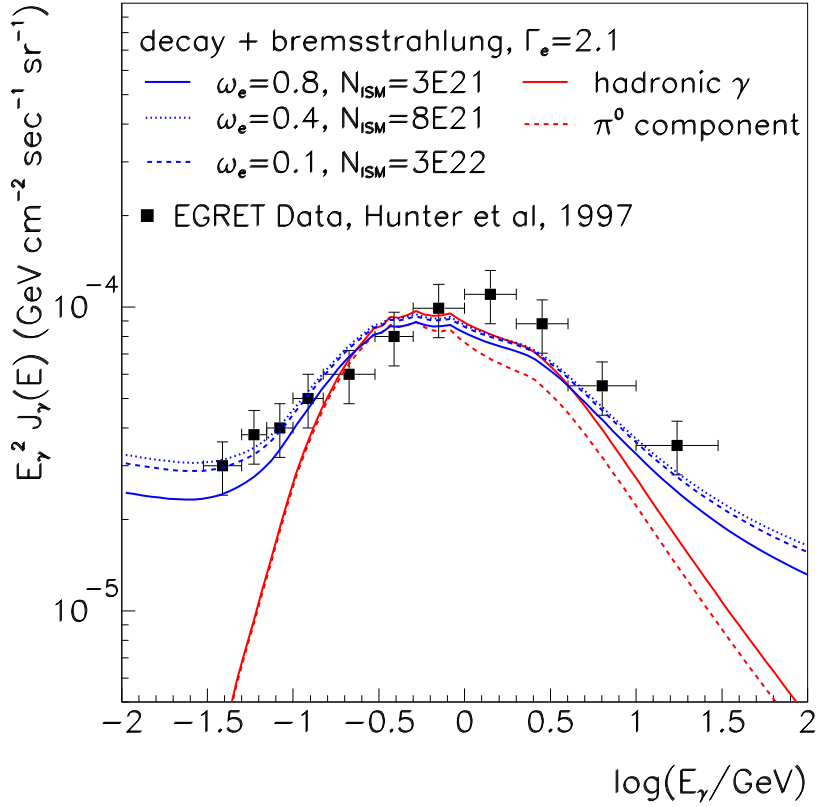}
\caption{Diffuse $\gamma$-ray spectrum at GeV range, shown in comparison with data for the inner part of Galaxy at $315^0 \le l \le 
345^0$ and $|b|\le 5^0$ \cite{Hunter97}. The model spectrum is composed of a
hadronic component calculated with cosmic-ray spectrum according to Eq.~(\ref{Eq:CRSpectraCutoff}),
plus a power-law component with spectral index $\Gamma_e=2.1$ for bremsstrahlung. 
Different curves correspond to different normalizations of the hadronic and 
leptonic contributions, here by varying the 
electron energy density $\omega_e=0.1,~0.4,~0.8~\textrm{eV/cm}^3$ 
and the gas column density 
$N_{ISM}=3~\times 10^{22},~8~\times 10^{21},~3~\times 10^{21}\textrm{cm}^{-2}$. 
The energy density of hadronic cosmic rays is kept constant at $\rho_E=0.75~\textrm{eV/cm}^3$.}\label{Fig:GeVBump}
\end{figure}
\end{center}

Figure~\ref{Fig:GeVBump} shows the observed GeV-band $\gamma$-ray emission from the inner Galaxy
in comparison with the contributions from $\pi^0$ decay as well as bremsstrahlung emission,
which we here describe by a power-law spectrum $\Phi_{\textrm{B}}(E)$ such that
\begin{eqnarray}\label{EQ:BremssPowerLaw}
\Phi_{\textrm{B}}(E)\simeq 1.3 \times10^{-8} \frac{\omega_e}{0.1~\textrm{eV/cm}^3}
                              \cdot \frac{N_{ISM}}{10^{22}~\textrm{cm}^{-2}}
                              \cdot \left(\frac{E}{100~\textrm{MeV}}\right)^{2.0-\Gamma_e} \quad 
\frac{\textrm{erg}}{\textrm{cm}^2~\textrm{sec}~\textrm{sr}}
\end{eqnarray}
with power-law spectral index $\Gamma_e=2.1$, the electron energy density 
$\omega_e=0.1,~0.4,~0.8~\textrm{eV/cm}^3$, and the gas column density 
$N_{ISM}=3~\times 10^{22},~8~\times 10^{21},~3~\times 10^{21}\textrm{cm}^{-2}$, respectively. 
The generating cosmic-rays are assumed with an energy 
density $\rho_E=0.75~\textrm{eV/cm}^3$. 
Models based on the locally observed cosmic-ray spectra generally
predict a softer spectrum for the leptonic components, even after accounting for inverse 
Compton emission \cite{Hunter97}, so we may in fact overestimate the GeV-band intensity 
of the leptonic contribution.
Nevertheless, it is clearly seen in this figure, that in the total intensity an over-shooting around 
$E_{\gamma} \simeq ~\textrm{300-600~MeV}$ appears in the modelled $\gamma$-ray energy 
distribution, whereas a deficit is present above 1~GeV. The observed spectrum of diffuse emission is 
always harder than the model spectrum, and we therefore conclude that an inaccurate description of 
hadronic $\gamma$-rays is ruled out as the origin of the GeV excess, in line with earlier work 
\cite{Mori97}.

\section{The TeV-band spectrum of RX~J1713-3946}
As SNR's are believed to be the main sources of Galactic cosmic rays, it is worth
testing whether or not the observed $\gamma$-ray spectra from SNR can be well modelled in terms of 
hadronic interactions. The $\gamma$-ray production matrix presented in this paper is
a valuable tool in such an analysis.

The HESS collaboration has measured the TeV-band gamma-ray spectrum of the shell-type SNR
RX~J1713.7-3946 over more than two decades in energy and in approximately ten angular resolution
elements per diameter \cite{Aharonian06AA}.
The data we use here consist of 25 independent spectral flux measurement for the entire remnant, 
taken with the complete 
HESS array in 2004. Since SNR RX~J1713.7-3946
appears closely associated with dense molecular clouds along the line of sight \cite{Fukui03,Moriguchi05}, 
it is reasonable to expect a strong signal of hadronic $\gamma$-rays.

Given the integrated flux in the TeV band,
for the hadronic scenario the required total energy in accelerated cosmic-ray nucleons is
\begin{equation}
W_p^{\rm tot}\approx (2\cdot 10^{50}\ {\rm erg})\,\left({{D}\over {\rm kpc}}\right)^2
\,\left({{n}\over {\rm atoms/cm^{-3}}}\right)^{-1}
\end{equation}
There is kinematic evidence that the distance $D$ is $1.3\pm 0.4$~kpc \cite{Fukui03}, whereas
the weakness of thermal X-ray emission implies a density of $n\lesssim 0.05\ {\rm atoms/cm^{-3}}$ in 
the center
of the remnant \cite{Pannuti03} and a postshock density of $n\lesssim 0.02\ {\rm atoms/cm^{-3}}$
\cite{CassamChenai04}. The low postshock density indicates that about 1 solar mass or less
have been swept up by the SNR, which therefore should just be in transition from the adiabatic 
phase to the
Taylor-Sedov phase, when its ability to accelerate particles is presumable at its peak.
The low density also implies a total energy in cosmic rays of about $10^{52}$~erg, somewhat larger 
than the canonical value for the total kinetic energy in a supernova explosion.

A high energy density of energetic particles is in line with the notion of Fermi-type acceleration
at the forward shock of SNR. 
This process is intrinsically efficient,
and a significant fraction of the pressure and energy density will be carried by energetic particles.
In that case the shocks should be strongly modified,
because the energetic particles have a smaller adiabatic index and a much larger mean free path 
for scattering than does the quasi-thermal plasma. In addition the particles at the highest 
energy escape, thus making the shock essentially radiative and increasing the compression ratio
\cite{Blandford87}. 

While the details of the published acceleration
models differ, they essentially all predict a continuous hardening of the 
particle spectrum with increasing energy, until a cut-off occurs at an energy that is difficult to predict
\cite{Berezhko99}.
At particle energies in excess of 1~TeV, i.e. relevant for $\gamma$-ray production above 200~GeV,
the spectrum should be harder than $E^{-2}$ and display a positive curvature, unless the cut-off has 
already set in \cite{Amato06}.
We therefore chose to parametrize the spectrum of accelerated hadrons in RX~J1713.7-3946 as
\begin{equation}
N(E)=N_0\,\left({E\over {E_0}}\right)^{-s+\sigma\,\ln {E\over {E_0}}}\,
\Theta\left[E_{\rm max}-E\right]
\label{pspec}
\end{equation}
where $\Theta$ is the step function and
$E_0=15$~TeV is a normalization chosen to render variations in the power-law index $s$ 
statistically independent from the choice of spectral curvature, $\sigma$. The
cut-off energy, $E_{\rm max}$, is a free parameter, but should be $\gtrsim 1$~PeV, if SNR are to produce 
Galactic cosmic rays up to the knee at a few PeV, because RX~J1713.7-3946 appears to already 
turn into the Sedov phase. For cosmic-ray modified shocks we would expect $s < 2$ and $\sigma > 0$.
The normalization, $N_0$, is not a parameter in our analysis, because we normalize both the data and the model
to their value at 0.97~TeV, while propagating the flux error at that energy to the errors in the
normalized flux at all other energies. 

\begin{center}
\begin{figure}
\includegraphics[scale=0.7]{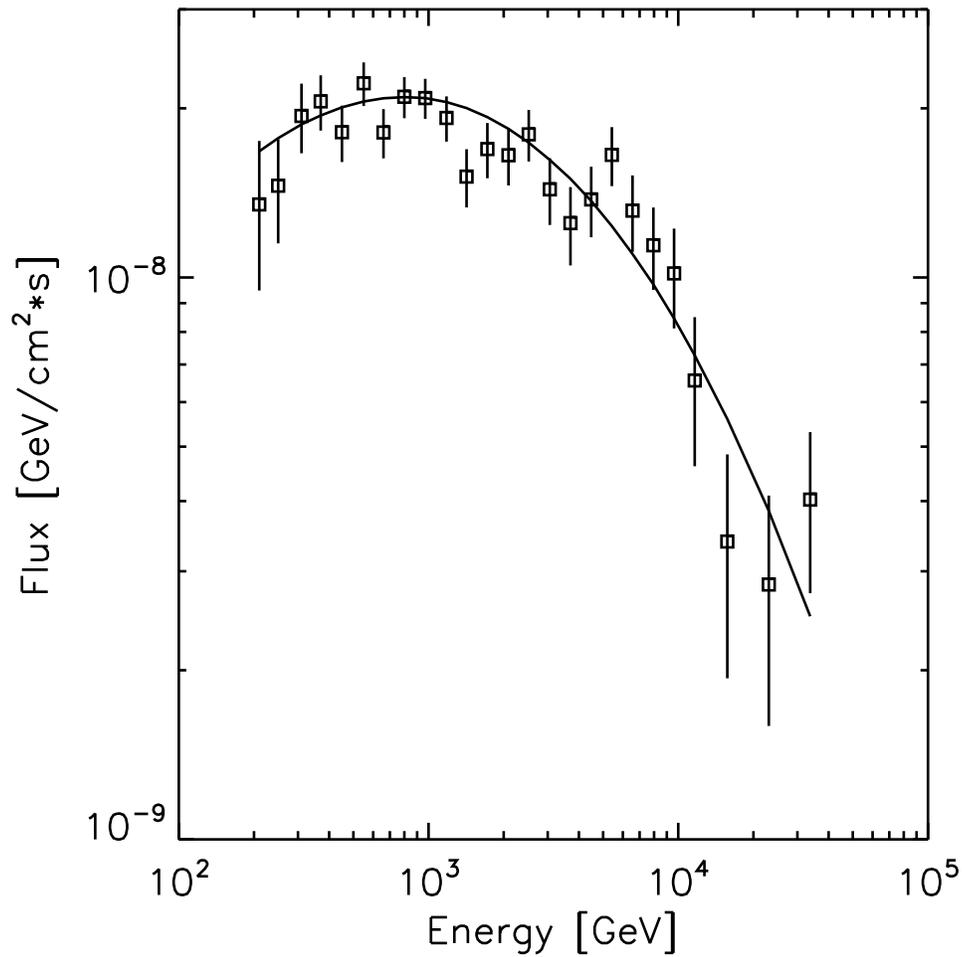}
\caption{The TeV-band $\gamma$-ray spectrum observed from RX~J1713-3946 with HESS, shown in comparison
with the best-fit model of hadronic gamma-ray production. Note that both data and model are normalized at 
0.97~TeV. Data in the energy range between
1~GeV and 200~GeV would be very valuable and may be provided by GLAST in the near future.}\label{Fig:BestFit}
\end{figure}
\end{center}

The parametrized particle spectrum Eq.~(\ref{pspec}) is used to calculate the $\gamma$-ray source function 
according to Eq.~(\ref{Eq:ProductionMatrix}), which then can be compared to the HESS data.
We use the $\chi^2$ statistic to both determine the goodness-of-fit and the confidence ranges of the three
parameters, $E_{\rm max}$, $s$, and $\sigma$. The best fit has a good quality with $\chi_{\rm min}^2=22.48$
for 22 degrees of freedom and is achieved for $s=2.13$, $\sigma=-0.25$, and $E_{\rm max}\gtrsim 200$~TeV,
i.e. no cut-off. The best fit is thus not commensurate with expectations based on acceleration at a
cosmic-ray modified shock. Figure~\ref{Fig:BestFit} shows the data in comparison with the best-fit model of hadronic
$\gamma$-ray production. To be noted from the figure is the obvious value of data in the energy range between
1~GeV and 200~GeV as may be provided by GLAST in the near future. The presently available upper limits
from EGRET \cite{Reimer02} are too high to provide meaningful constraints and are therefore omitted here.

\begin{center}
\begin{figure}
\includegraphics[scale=0.75]{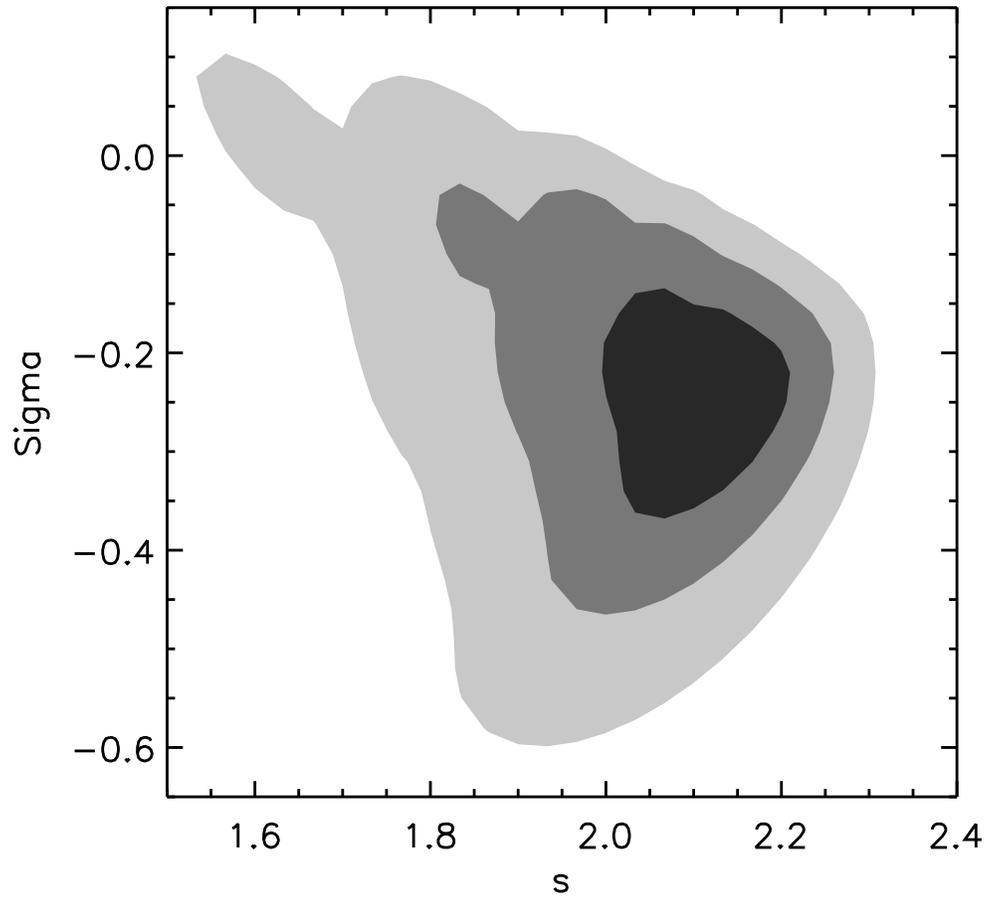}
\caption{The confidence regions for the spectral curvature $\sigma$ and the spectral index $s$. 
The dark shaded area corresponds to 68\% probability, or 1 sigma, the medium gray area indicates
95\% confidence, or 2 sigma, and light gray is for 99.7\% probability or 3 sigma. 
}\label{Fig:Sigma.S}
\end{figure}
\end{center}

\begin{figure}
\begin{center}
\includegraphics[scale=0.8]{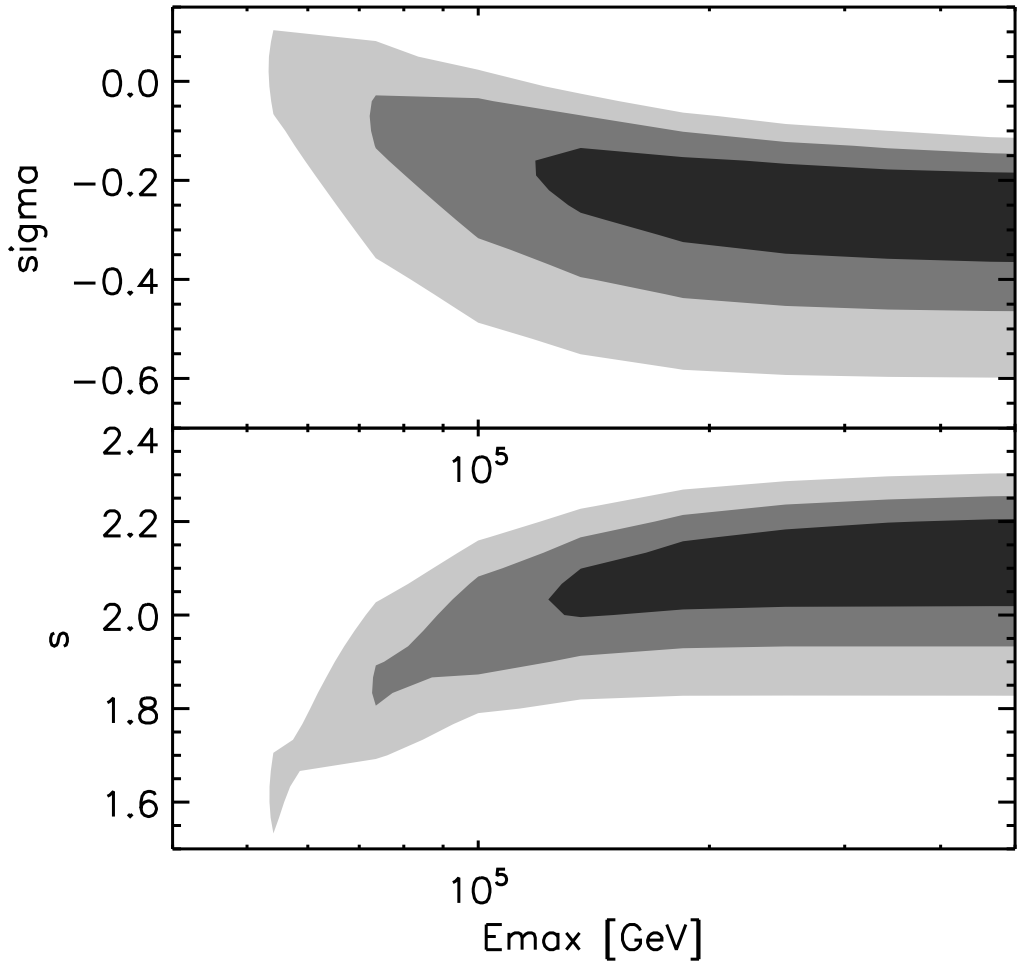}
\caption{Upper: The confidence regions for the spectral curvature $\sigma$ and the spectral index $s$, 
versus the cut-off energy $E_{\rm max}$. The contour levels are 1 sigma, 2 sigma, and 3 sigma as in 
Figure~\ref{Fig:Sigma.S}.}\label{Fig:Sigma.S.Emax}
\end{center}
\end{figure}

More important than the best fit are the confidence ranges of parameters, for which the 
probability distributions follow a $\chi_n^2$ distribution in  $\chi^2 - \chi_{\rm min}^2$,
where $n$ is the number of parameters, for which the combined confidence volume is determined
\cite{Lampton76}. As only two-dimensional contours can be unambiguously displayed, we have performed 
a 2-parameter analysis by marginalizing one parameter. The areas for the three confidence levels 
68\%, 95\%, and 99.7\%, corresponding to 1, 2, and 3 sigma, respectively, are shown in 
Figure~\ref{Fig:Sigma.S} for the curvature $\sigma$ versus the power index $s$, and in Figure~\ref{Fig:Sigma.S.Emax} for
$\sigma$ and $s$ versus the cut-off energy $E_{\rm max}$.

To be noted from the figures is that a positive spectral curvature, $\sigma > 0$, is excluded at more
than 95\% confidence. This is clearly at odds with expectations based on cosmic-ray modified shocks.
The contour for $E_{\rm max}$ are open toward higher energies, indicating that a cut-off is not
statistically required. A very high $E_{\rm max}$ would be in line with cosmic-ray acceleration to the knee
at a few PeV, but one should note from Figure~\ref{Fig:Sigma.S.Emax} that in that case
a negative spectral curvature is strongly required, in fact $\sigma < -0.1$ with more than 99.7\%
confidence! A small or marginally positive curvature is permitted only in combination with a
low cut-off energy $E_{\rm max}\lesssim 100$~TeV. 

These statements are verified by additional tests in which we allowed only a positive spectral
curvature, $\sigma \ge 0$. The best fit under those constraints achieved $\chi^2=29.39$, higher than the 
unconstrained best fit by $\Delta\chi^2=6.91$, which again demonstrates how much poorer the fit is for
$\sigma \ge 0$. The two parameters $E_{\rm max}$ and $s$ would be slightly correlated with pivot point
at $s\simeq 1.8$ and $E_{\rm max}\simeq 75$~TeV, so a lower cut-off energy would compensate for a harder 
power index and vice versa. Nevertheless the cut-off energy would be $E_{\rm max}< 140$~TeV
with 99.7\% confidence. 

These findings clearly demonstrate that a softening particle spectrum is statistically required, if the 
TeV-band $\gamma$-ray spectrum from RX~J1713.7-3946 is of hadronic origin. A more gradual decay of the spectrum
is preferred over a hard cut-off. With the best-fit curvature $\sigma=-0.25$ the particle spectral index
would change by $\Delta s=0.58$ for each decade in energy, or by $\Delta s=1.04$ between our normalization energy
$E_0=15$~TeV and 1~PeV. There is no evidence for either efficient nucleon
acceleration to a energies near the location of the knee, or for nonlinear acceleration at
cosmic-ray modified shocks. 

\section{Conclusion}
The diffuse galactic $\gamma$-ray emission provides information to obtain a proper understanding 
of the origin of galactic cosmic rays. In this work, 
the $\gamma$-ray spectrum generated in cosmic-ray interactions with ISM gas is calculated using the simulation code 
DPMJET-III. For the first time we consider both protons and helium nuclei in cosmic 
rays and include all secondary products with $\gamma$-rays as 
the final decay state as well as direct $\gamma$-ray production and the intermediate resonance production. We find that about 20\% of 
the $\gamma$-ray spectrum come from non-$\pi^0$ secondaries, to which directly produced 
$\gamma$-ray photons contribute stronger than other secondaries. We also calculate the energy loss rate
of hadronic cosmic rays in the Galaxy and present a simple analytical approximation to it, that may be used in
cosmic-ray propagation studies.

The main product of this work are matrices, or look-up tables, that can be used to calculate the source spectra
of $\gamma$-rays and other secondary products arising from inelastic collisions of cosmic rays with arbitrary spectrum.
Our study involves $\gamma$-rays with energies up to $10^6$~GeV and incoming cosmic rays up to $10^8$~GeV. 
The matrices will be made available for download at {\it http://cherenkov.physics.iastate.edu/gamma-prod}.

We apply the $\gamma$-ray production matrices to two examples: the GeV excess and the TeV-band $\gamma$-ray
spectrum of the shell-type SNR RX J1713-3946. Our findings are:
\begin{itemize}
\item The modifications in the GeV-band $\gamma$-ray emission of hadronic origin are insufficient to 
explain the GeV excess in diffuse galactic $\gamma$-rays. It appears that, to arrive at an understanding
of the GeV excess, GLAST data are needed to better understand 
its angular distribution and exact spectral form.
\item If the TeV-band spectrum of RX J1713-3946 as observed with HESS is caused by cosmic-ray nucleons, then 
a soft cut-off at about 100~TeV is statistically required in the particle spectrum. 
There is no evidence for efficient nucleon acceleration to energies near the knee in the cosmic-ray spectrum.
We also find no evidence of the spectral curvature and hardness predicted by standard models
of cosmic-ray modified shock acceleration, even though a strong cosmic-ray modification must be expected given
the high cosmic-ray density in the remnant implied by the low thermal gas density and high $\gamma$-ray luminosity.
We again emphasize the need for GLAST data to better constrain the $\gamma$-ray spectrum below 100~GeV. 
\end{itemize}

\acknowledgments
The author C.-Y. Huang would like to thank R. Engel and M. Bu\'enerd for the helpful discussions. 
Support by NASA under award No. NAG5-13559 is gratefully acknowledged.

%
%
\clearpage

\clearpage

\clearpage

\clearpage

\clearpage

\clearpage

\clearpage

\clearpage


\clearpage




\clearpage

\begin{thebibliography}{99}
\bibitem{Nishimura80}
J. Nishimura et al., 
\APJ{238}{1980}{394}
\bibitem{Erlykin98} 
A. D. Erlykin, M. Lipski and A. W. Wolfendale, 
\AstroPartPhys{8}{1998}{283}
\bibitem{Cowsik79} 
R. Cowsik and  M. A. Lee, 
\APJ{228}{1979}{297}
\bibitem{Pohl98}
M. Pohl and J. Esposito,
\APJ{507}{1998}{327}
\bibitem{Buesching05}
I. B\"usching et al.,
\APJ{619}{2005}{314}
\bibitem{Hunter97}
S. D. Hunter et al.,
\APJ{481}{1997}{205}
\bibitem{Aharonian00} 
F. A. Aharonian and A. M. Atoyan, 
\AA{362}{2000}{937}
\bibitem{Dermer86APJ} 
C. D. Dermer, 
\APJ{307}{1986}{47}
\bibitem{Mori97} 
M. Mori, 
\APJ{478}{1997}{225}
\bibitem{Stephens81} 
S. A. Stephens and G. D. Badhwar, 
\APSS{76}{1981}{213}
\bibitem{Pohl02} 
M. Pohl, 
Proc. The Universe Viewed in $\gamma$-rays, 
Froniter Science Service No. 39, Universal Academy Press, 2003
\bibitem{Strong04}
A. W. Strong, I. V. Moskalenko and O. Reimer,
\APJ{613}{2004}{962}
\bibitem{Pohl00}
M. Pohl and R. Schlickeiser,
\AA{354}{2000}{395} 
\bibitem{Muecke01}
A. M\"ucke and R. J. Protheroe, 
\AstroPartPhys{15}{2001}{121}
\bibitem{Muecke03}
A. M\"ucke, R. J. Protheroe, R. Engel, J. P. Rachen and T. Stanev,   
\AstroPartPhys{18}{2003}{593}
\bibitem{Boettcher04}
M. B\"ottcher and A. Reimer, 
\APJ{609}{2004}{576}
\bibitem{Aharonian04Nature}
The HESS Collaboration, F. A. Aharonian et al., 
\Nature{432}{2004}{75}
\bibitem{Aharonian05AA}
The HESS Collaboration, F. A. Aharonian et al., 
\AA{437}{2005}{L7}
\bibitem{Aharonian06AA}
The HESS Collaboration, F. A. Aharonian et al., 
\AA{449}{2006}{223}
\bibitem{Huang05ICRC}
C.-Y. Huang, 
29th Int. Conf. Cosmic Rays 4 (2005) 73.
\bibitem{Blattnig00} 
Steve R. Blattnig, Sudha R. Swaminathan, Adam T. Kruger, Moussa Ngom and John W. Norbury, 
\PRD{62}{2000}{094030}
\bibitem{DomingoSantamaria05}
E. Domingo-Santamar\l{i}a and D. F. Torres, 
\AA{444}{2005}{403}
\bibitem{Kamae05}
T. Kamae, A. Toshinori and K., Tatsumi,
\APJ{620}{2005}{244}
\bibitem{Kamae06}
T. Kamae, N. Karlsson, M. Tsunefumi, A. Toshinori, K., Tatsumi,
\APJ{647}{2006}{692}
\bibitem{Kelner06}
S. R. Kelner, F. A. Aharonian and V. V. Bugayov,
\PRD{74}{2006}{034018}
\bibitem{Gaisser90}
T. K. Gaisser, 
Cosmic Rays and Particle Physics, Cambridge University Press, 2000
\bibitem{Roesler00}
S. Roesler, R. Engel and J. Ranft, 
Advanced Monte Carlo for Radiation Physics, Particle Transport Simulation and Applications (MC 2000), Lisbon, 2000
\bibitem{Lipkin72} 
H. J. Lipkin and M. Peshkin, 
\PRL{28}{1972}{862}
\bibitem{Jaeger75I} 
K. Jaeger et al., 
\PRD{11}{1975}{1756}
\bibitem{Jaeger75II} 
K. Jaeger et al., 
\PRD{11}{1975}{2405}
\bibitem{Buesser76} 
F. W. B\"usser et al., 
\NPB{106}{1976}{1}
\bibitem{Liu03} 
Y. Liu, L. Derome and M. Bu\'enerd, 
\PRD{67}{2003}{073022}
\bibitem{Dermer86AA} 
C. D. Dermer, 
\AA{157}{1986}{223}
\bibitem{Tan83} 
L. C. Tan and L. K. Ng, 
\JPhysG{9}{1983}{227}
\bibitem{Stecker70} 
F. W. Stecker, 
\APSS{6}{1970}{377}
\bibitem{Ranft95}
J. Ranft,
\PRD{51}{1995}{64}
\bibitem{Knapp03}
J. Knapp et al., 
\AstroPartPhys{19}{2003}{77}
\bibitem{Kasahara02}
K. Kasahara et al., 
\PRD{66}{2002}{052004}
\bibitem{Kass79} 
R. D. Kass et al., 
\PRD{20}{1979}{605}
\bibitem{Pohl03}
M. Pohl, C. Perrot, I. Grenier and S. Digel, \AA{409}{2003}{581}
\bibitem{Fukui03}
Y. Fukui et al., \PASJ{55}{2003}{L61}
\bibitem{Moriguchi05}
Y. Moriguchi et al., \APJ{631}{2005}{947}
\bibitem{Pannuti03}
T. G. Pannuti, G. E. Allen, J. C. Houck and S. J. Sturner, \APJ{593}{2003}{377} 
\bibitem{CassamChenai04}
G. Cassam-Chena\"i et al., \AA{427}{2004}{199}
\bibitem{Blandford87}
R. D. Blandford and D. Eichler, \PhysRept{154}{1987}{1}
\bibitem{Berezhko99}
E. G. Berezhko and D. C. Ellison, \APJ{526}{1999}{385}
\bibitem{Amato06}
E. Amato and P. Blasi, \MNRAS{371}{2006}{1251} 
\bibitem{Reimer02}
O. Reimer and M. Pohl, \AA{390}{2002}{L43}
\bibitem{Lampton76}
M. Lampton, B. Margon and S. Bowyer, \APJ{208}{1976}{177}
\end{thebibliography}
\end{document}